\documentclass[twocolumn,showpacs,preprintnumbers,amsmath,amssymb]{revtex4}

\usepackage{pdfpages}
\usepackage{graphicx}% Include figure files
\usepackage{dcolumn}% Align table columns on decimal point
\usepackage{bm}% bold math
\usepackage{extarrows}
\usepackage{lipsum}
\usepackage{color}
\usepackage{chemfig}
\usepackage{comment}
\usepackage{enumitem}

\begin{document}

%\preprint{APS/123-QED}

\title{Insights into the relation between noise and biological complexity}
% Force line breaks with \\

\author{Fabrizio Pucci}\email{fapucci@ulb.ac.be}
\author{Marianne Rooman} \email{mrooman@ulb.ac.be}
 \affiliation{Department of BioModeling, BioInformatics \& BioProcesses, Universit\'e Libre de Bruxelles,
Roosevelt Ave. 50, 1050 Brussels, Belgium}

\date{\today}% It is always \today, today,

\begin{abstract}
Understanding under which conditions the increase of systems complexity is evolutionary advantageous, and how this trend is related to the modulation of the intrinsic noise, are fascinating issues of utmost importance for synthetic and systems   biology. To get insights into these matters, we  analyzed  chemical reaction networks with different topologies and degrees of complexity,  interacting or not with the environment. We showed that the global level of fluctuations at the steady state, as measured by the sum of the Fano factors of the number of molecules of all species, is directly related to the topology of the network. For systems with zero deficiency, this sum  is constant and equal to the rank of the network. For higher deficiencies, we  observed an increase or decrease of the fluctuation levels according to the values of the reaction fluxes that link internal species, multiplied by the associated stoichiometry. We showed that the noise is reduced when the fluxes all flow towards the species of higher complexity,  whereas it is amplified when the fluxes are directed towards lower complexity species. 
\end{abstract}


\maketitle
\section{Introduction}

Fluctuations play a major role in the dynamics of a large variety of complex biological processes. To properly describe the stochastic and heterogeneous nature  of systems such as  the transcriptional machinery \cite{Rev2} or cell differentiation processes \cite{Rev3}, stochastic modeling approaches are indispensable  \cite{Rev1}. For example, intensive efforts have been devoted in the last decade to the characterization of stochasticity in chemical reaction networks (CRNs) by studying the propagation of fluctuations through the networks  \cite{AndersonI,AndersonII} or by analyzing the relation between thermodynamic properties and noise levels \cite{Polettini,Rao}. However, the full understanding of the stochastic properties of biomolecular networks  remains an intricate  goal that is far from being  met.

A challenging open question concerns the relationship between the complexity of biological systems and the modulation of the intrinsic noise.  At first glance, the  evolutionary pressure, which led from  unicellular organisms to higher eukaryotes, tends to favor both complexity increase and noise reduction, but this is clearly  not a general rule. On the basis of modeling investigations, it has been shown that, for some systems,  the increase of systems complexity is  associated to a decrease of the noise level \cite{Cardelli}, whereas other   processes are  noise-driven \cite{Hoffmann}; for still other systems, the intrinsic noise can  be either increased or decreased according to the  parameter values of the model \cite{MarianneRooman}. 

The lack of a general understanding is due to the fact that CRNs describing biological systems are usually very large and complex and thus  complicated to  model mathematically, especially via stochastic simulations in which  the parameter space becomes rapidly too large to be tractable. Hence, only toy models can be realistically analyzed. 

 The purpose of this work is to deepen our knowledge on noise modulation in stochastic CRNs. This was done by analyzing model systems with different degrees of complexity and by exploring analytically and numerically their dynamical behavior and  parameter spaces, using  the It\=o stochastic differential equations formalism. 
For the systems studied, we  observed general relations between some structural characteristics of the CRNs and the noise levels evaluated by the Fano factors of the biochemical species involved. We conjecture the validity of these relations for general classes of CRNs.

\section{Stochastic Chemical Reaction Networks}

Let us briefly review some basic concepts of chemical reaction networks. They are defined by the  $\left[\mathcal{S},\mathcal{C},\mathcal{R}\right]$ triplets consisting of  sets of chemical species $\mathcal{S}$, complexes $\mathcal{C}$ and elementary reactions $\mathcal{R}$. In the case of open systems, the environment is  considered as a complex with vanishing stoichiometric coefficients. Let $U_i(t)$ denote the  number of molecules  of species $i$ (with $1 \le i \le $card($\mathcal{S}$)) at time $t$. It  satisfies the  chemical Langevin equation (CLE) in the It\=o formalism \cite{CLE}:
\begin{equation}
\frac{dU_i(t)}{dt} = \sum_{j=1}^{\text{card}(\mathcal{R})} k_{ij} a_j ( \textbf{U}(t)) + \sum_{j=1}^{\text{card}(\mathcal{R})} k_{ij}  \sqrt{a_j (\textbf{U}(t))}   \Gamma_j (t)
\label{uno}
\end{equation}
where $k_{ij}$ is the stoichiometry matrix, $ a_j ( \textbf{U}(t))$  the  rate of  reaction $j$  and $\Gamma_j (t)$  statistically independent Gaussian white noises (Wiener processes). This equation  describes the temporal evolution of $\textbf{U}(t)$  and implies that its conditioned probability density function obeys the associated Fokker-Planck equation. 
 Here we assume a mass-action kinetic scheme for the CRNs, which means that the rate of a chemical reaction is proportional to the product of the concentrations of the reactants raised to powers equal to their stoichiometric coefficients.

CRNs are said to be  complex balanced \cite{Feinber01,Feinber02} if, for each complex $y$, the sum of the mean reaction rates for the reactions $r \subset \mathcal{R}$ for which $y$ is a reactant complex is equal to  the sum of the mean reaction rates for $r' \subset \mathcal{R}$ for which $y$ is a product complex at the steady state $U$: 

\begin{equation}
\sum_{j\in r} \textbf{E}(a_j( \textbf{U}))  = \sum_{j \in r'} \textbf{E}(a_j( \textbf{U})). 
\label{due}
\end{equation}
Detailed balanced CRNs are a subclass of complex balanced systems for which this relation holds separately for each pair of forward and inverse reactions linking  two complexes. Detailed balanced steady states correspond to thermodynamic equilibrium states, whereas  the others are  non-equilibrium steady states (NESS).

To characterize the topology of a given CRN, the notion of deficiency has been introduced \cite{Feinber01,Feinber02} as 

\begin{equation}\delta = \text{card}(\mathcal{C})- L -\mathcal{X},
\end{equation}

\noindent
where $L$  is the number of linkage classes, namely the number of  connected components of the reaction network, and $ \mathcal{X}$ is the dimension of the stoichiometry subspace,  namely the rank of the network. 

CRNs are also characterized by their reversibility properties.  They are said to be  reversible if the existence of a reaction that turns one complex into another implies the existence of the reverse reaction. They are  weakly reversible if the existence of a reaction path from  complex one to complex two implies the existence of a path from  complex two to complex one. 

We focused here on open and closed, weakly reversible, mass-action CRNs of arbitrary deficiency, which admit a steady state that is not always complex balanced.

\section{Model Systems}

We studied three types of weakly reversible CRNs, depicted in  Fig.\ref{fig1}. The simplest one (Fig.\ref{fig1}a)  is defined by the reaction chain: 
\begin{equation} 
\schemestart
   $n$\, X \arrow{<=>}[,.7,line width=.4pt] Y\arrow{<=>}[,.6,line width=.4pt]$\varnothing$\arrow{<=>}[,.6,line width=.4pt] X
\schemestop
\label{CRN1}
\end{equation}
It  represents for example the assembly of monomeric molecules X into homo-oligomers Y, where $n \in  \mathbb{N}_{>0}$ indicates the number of monomers in each oligomer. When $n=1$, this CRN represents the interconversion between two states of the same molecule ({\it e.g.} activated or not) or between two localizations ({\it e.g.} intra- or extracellular).  Both species are linked to the environment. 

The  system of It\=o stochastic differential equations (SDE) that describes this CRN reads as (see Eq. (\ref{uno})): 
\begin{eqnarray}
dX(t) &=& dP_X(t) - dD_X(t) - n\, (dF(t)- dG(t)) \nonumber \\
dY(t) &=& dP_Y(t) - dD_Y(t)+dF(t)- dG(t) 
\label{oligomerization0}
\end{eqnarray}
where $X(t)$ and $Y(t)$ indicate the number of molecules of species X and Y. $dP_X(t)$ and $dP_Y(t)$ are the production terms for the variables $X(t)$ and $Y(t)$ respectively,  $dD_X(t)$ and $dD_Y(t)$ are the associated degradation terms, while $dF(t)$ and $dG(t)$ are  interconversion terms. Assuming mass-action kinetics, these terms can be written as: 
 
\begin{eqnarray}
dP_X(t) &=& p_x dt + \sqrt{p_x}\, \Gamma_{p_x} (t) \nonumber \\
dD_X(t) &=& d_x X(t) dt  +  \sqrt{d_x X(t)}\, \Gamma_{d_x} (t) \nonumber \\
dP_Y(t) &=& p_y dt + \sqrt{p_y}\, \Gamma_{p_y} (t) \nonumber \\
dD_Y(t) &=& d_y X(t) dt  +  \sqrt{d_y X(t)}\, \Gamma_{d_y} (t) \nonumber \\
dF(t)  &=& f X(t)^n dt +  \sqrt{f X(t)^n}\, \Gamma_{f} (t) \nonumber \\
dG(t) &=& g Y(t) dt + \sqrt{g Y(t)}\, \Gamma_{g} (t) 
\label{oligomerization1}
\end{eqnarray}
where $p_x, d_x, p_y, d_y, f,g \in \mathbb{R}_{\ge 0}$  are the parameters of the model. We assumed here that the number of molecules is large enough so that the approximation $X^n\approx X(X-1)...(X-n+1)$ holds.

We also studied  more complex CRNs, which model for instance a biological system in which a molecular species undergoes  a homo-oligomerization process through an intermediate oligomerization step of lower order (Fig.\ref{fig1}b):
\begin{eqnarray}  
\schemestart
  $n$\, X \arrow{<=>}[,.7,line width=.4pt] Y$\quad$,  $\quad m$\, Y \arrow{<=>}[,.7,line width=.4pt] Z
\schemestop\nonumber \\
 \schemestart
  X \arrow{<=>}[,.6,line width=.4pt]$\varnothing \quad$,  $\quad $ Y \arrow{<=>}[,.6,line width=.4pt]$\varnothing \quad$,  $\quad$ Z\arrow{<=>}[,.6,line width=.4pt]$\varnothing$
\schemestop
\label{CRN2}
\end{eqnarray}
with $n, m \in  \mathbb{N}_{>0}$, or  in which the homo-oligomerization process occurs with or without an intermediate step (Fig.\ref{fig1}c):
\begin{eqnarray}  
\schemestart
  $n$\, X \arrow{<=>}[,.7,line width=.4pt] Y$\;$,  $\; m$\, Y \arrow{<=>}[,.7,line width=.4pt] Z $\;$, $\; n m$ X \arrow{<=>}[,.7,line width=.4pt] Z
\schemestop\nonumber \\
 \schemestart
  X\arrow{<=>}[,.6,line width=.4pt]$\varnothing \;$,  $\;$ Y\arrow{<=>}[,.6,line width=.4pt]$\varnothing \;$,  $\;$ Z\arrow{<=>}[,.6,line width=.4pt]$\varnothing$
\schemestop
\label{CRN3}
\end{eqnarray}
The SDEs modeling these CRNs can be easily obtained by generalizing  Eqs. (\ref{oligomerization0}-\ref{oligomerization1}).

 \section{Intrinsic Noise}

A central parameter that quantifies the role of fluctuations in  biochemical systems is the Fano factor $F(t)$,  defined as the ratio between the variance of a stochastic variable $U(t)$ and its mean: $F(t)= \textbf{Var}[U(t)]/  \textbf{E}[U(t)]$. If the variable follows a Poisson distribution, its Fano factor $F$ is equal to one. When $F$ is larger than one, the intrinsic noise affects more strongly the  variable concentration, and the distribution is called super-Poissonian. The distribution is  sub-Poissonian when $F < 1$. 

To analyze the role of the fluctuations in the different types of CRNs depicted in Fig.\ref{fig1}, we  computed  the sum of the Fano factors of all  species at the steady state as a function of the system's parameters.  This Fano factor sum is taken to represent the global noise level of the system. We obtained its  expression from the SDEs  of Eqs (\ref{oligomerization0}-\ref{oligomerization1}) by employing an Euler-Mayurama time discretization scheme  \cite{EulerMayurama}. The equations were studied  analytically with a moment closure approximation \cite{Grima} as well as through numerical simulations. A detailed analysis of these systems will be presented in an upcoming paper \cite{PucciRoomanII}; here we  show the main results.

\subsection{Deficiency zero}
We studied a series of $\delta=0$ CRNs among which the following three classes :
\begin{enumerate}[label=(\alph*)]
\item Closed CRNs described in Figs \ref{fig1}a-b with $p_i=d_i=0$ for all species $i$; 
\item Open CRNs described in Figs \ref{fig1}a-b with $p_i=d_i=0$ for all but one species; 
\item Open CRNs described in Figs \ref{fig1}a-c with  $n=m=1$. 
\end{enumerate}
Types (a) and (b) are detailed balanced systems, admitting equilibrium states for all reaction rates, while (c) is only complex balanced. 

For all these CRNs we  found analytically  that the sum of  the steady state Fano factors over all species $i$  is equal to the rank :
\begin{equation}
\sum_{i=1}^{\text{card}(\mathcal{S})} F_i =  \mathcal{X} 
\label{marchetti} 
 \end{equation}  
It can be checked that all CRNs with deficiency zero satisfy this simple relation. It is directly related  to the result  shown for this kind of CRNs in \cite{Anderson}:  the steady state probability distribution of the number of molecules of each species  can be expressed as a product of Poisson distributions,  or as a multinomial distribution in the case some  conservation laws hold and thus the state  space is reduced, for example in closed systems where the number of molecules is conserved.

\subsection{Deficiency one}
For systems with deficiency larger than zero that admit a stationary solution, the situation becomes more complicated due to the correlation between the chemical species in the stationary state.  

\begin{figure}	
	\includegraphics[scale=0.42]{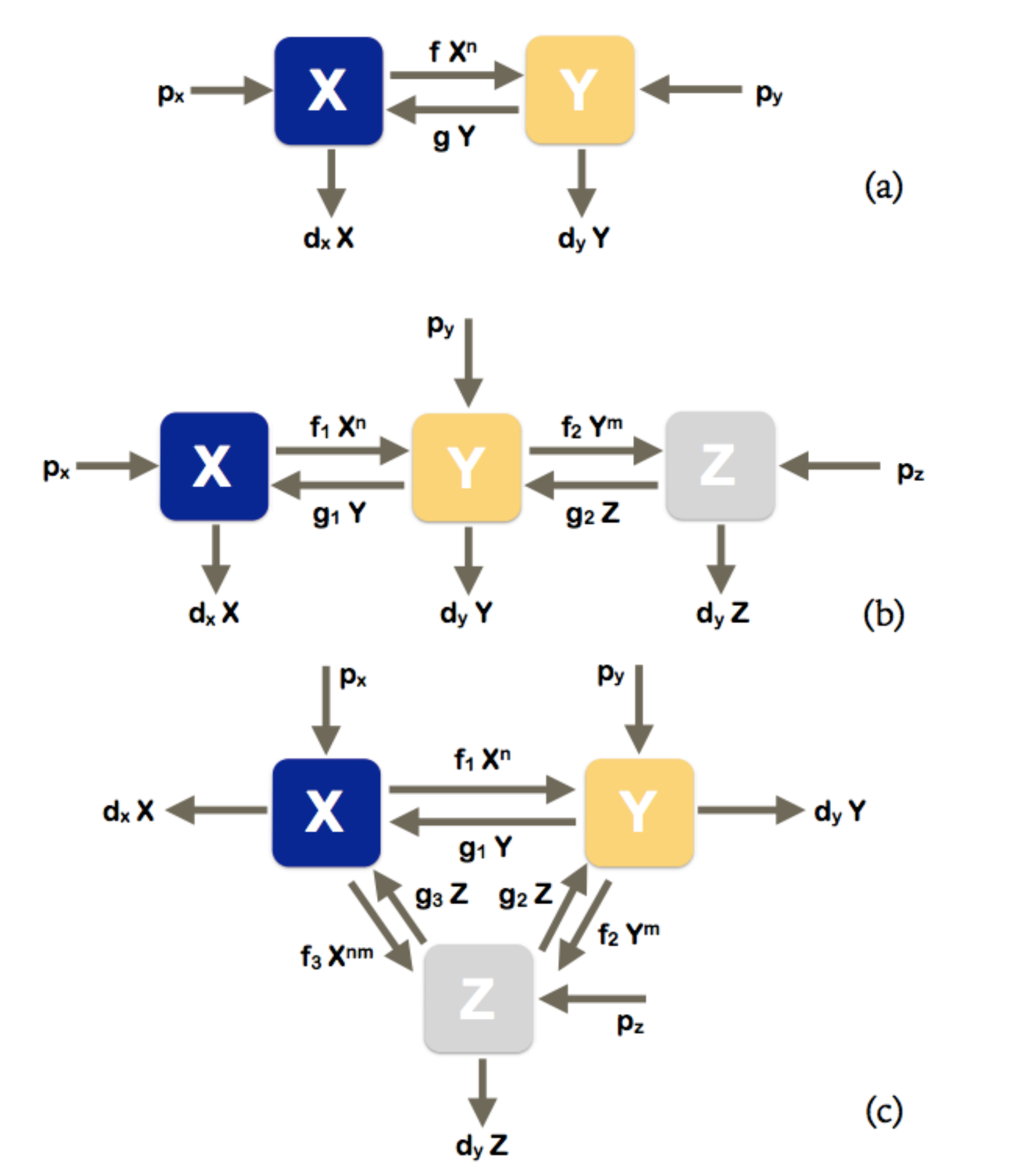}
	\caption{Schematic picture of the reaction networks analyzed in this letter.} 
\label{fig1}
\end{figure}

Consider first the $\delta=1$ homo-oligomerization CRN shown in Fig. \ref{fig1}a and Eq. (\ref{CRN1}) with $n>1$ and the two species X and Y linked to the environment. Note that in this model, both monomers and oligomers can be degraded, but also   produced; to obtain the more realistic case in which only the monomer is produced, the oligomer  production rate $p_y$ must be set to zero.

\begin{figure}	
	\includegraphics[scale=0.265]{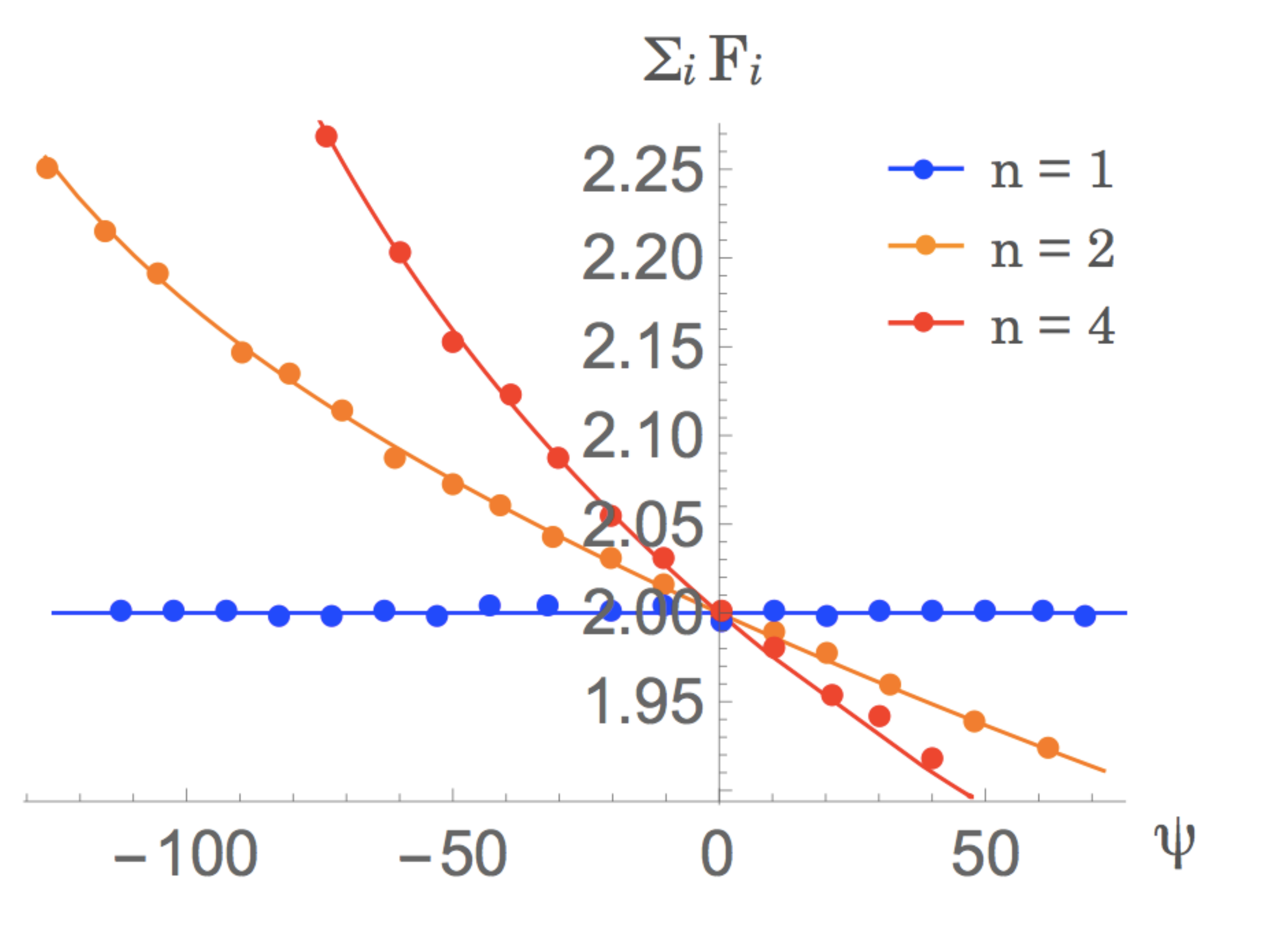}
\caption{Sum of Fano factors over the two different species as a function of the mean flux $\Psi$,  for different values of $n$ (1, 2 and 4) and  different values of the parameter $f$, for the SDEs shown in Fig. \ref{fig1}a; all other parameters are kept fixed: $p_x=p_y=200$, $d_x=d_y=0.001$ and $g=0.002$. The analytical solutions are indicated with lines and the solutions from stochastic simulations with dots.} 
\label{Fano}
\end{figure}

Using the same analytical procedure as for the $\delta=0$ case, we showed that the internal flux $\Psi$ that flows between the two species X and Y:
  \begin{equation}
         \Psi = f\, \textbf{E}[X^n] - g \,  \textbf{E}[Y]       
    \end{equation}
is generally non-zero. The other mean fluxes  of the system involve exchanges with the environment and are  proportional  to $\Psi$.
We  obtained the analytical relation for the sum of the Fano factors in terms of the rank and the mean flux:    
\begin{equation}
\sum_{i=1}^{\text{card}(\mathcal{S})} F_i =  \mathcal{X}  - \alpha  \,  (n-1) \,  \Psi
 \label{eq111}
 \end{equation}
where $\alpha$ is a positive function of the system's parameters.   
The rank  is here equal to $\mathcal{X}=2$. As expected, this equation reduces to Eq. (\ref{marchetti}) when $n=1$ ({\it i.e.} the system is complex balanced) or $\Psi=0$ ({\it i.e.} the steady state is detailed balanced).  The analytical and numerical results for the sum of Fano factors as a function of $\Psi$,  for different values of $n$ and  the parameter $f$, are plotted in Fig. \ref{Fano}.

Relation (\ref{eq111}) means that there is a global reduction of the intrinsic noise level, as measured by the sum of Fano factors, when the mean flux $\Psi$ is positive, thus when the  net flux is directed towards the chemical species with higher degree of complexity (here the oligomer Y).  When the flux  is directed towards the species with lower complexity (here the monomer X), there is a global amplification of the noise.  We would like to emphasize that the larger the complexity level, the larger the reduction or amplification effect.  Indeed, the proportionality coefficient between the Fano factor sum and the flux increases in absolute value with  $n$. 

This  relation  can easily be generalized to the other $\delta=1$ systems  depicted in Fig. \ref{fig1}b-c, for the parameter values for which there is only one independent mean flux linking two complexes of different reaction stoichiometry. Note that, in the case of several (dependent) fluxes, the positiveness of the proportionality function $\alpha$ is  only ensured if they all  flow from lower to higher complexity or $\emph{vice versa}$.

\subsection{Higher deficiencies}
To get insights into the intrinsic noise in more complex CRNs and generalize Eq. (\ref{eq111}), we  studied networks with higher  deficiency values.  First we considered the CRN of deficiency $\delta=2$ shown in Fig. \ref{fig1}b and Eq. (\ref{CRN2}), with $n,m>1$ and the three species X, Y and Z linked to the environment. It has two independent fluxes that flow between pairs of species: 
\begin{eqnarray}  
\Psi_{1}&=&  f_1\, \textbf{E}[X^n] - g_1 \,  \textbf{E}[Y] \nonumber \\
\Psi_{2}&=& f_2\, \textbf{E}[Y^m] - g_2 \,  \textbf{E}[Z] 
\label{flux}
\end{eqnarray}
The fluxes that link the species to the environment are linear combinations of these two fluxes. 

We also considered the more complex  case with deficiency $\delta=3$ shown in Fig. \ref{fig1}c and Eq. (\ref{CRN3}). For $n,m>1$ and all species linked to the environment, this system has three independent fluxes, the two given by Eq. (\ref{flux}) and a third one:
\begin{equation}  
\Psi_{3}=  f_3\, \textbf{E}[X^{nm}] - g_3 \,  \textbf{E}[Z]
\end{equation}

\begin{figure}	
	\includegraphics[scale=0.25]{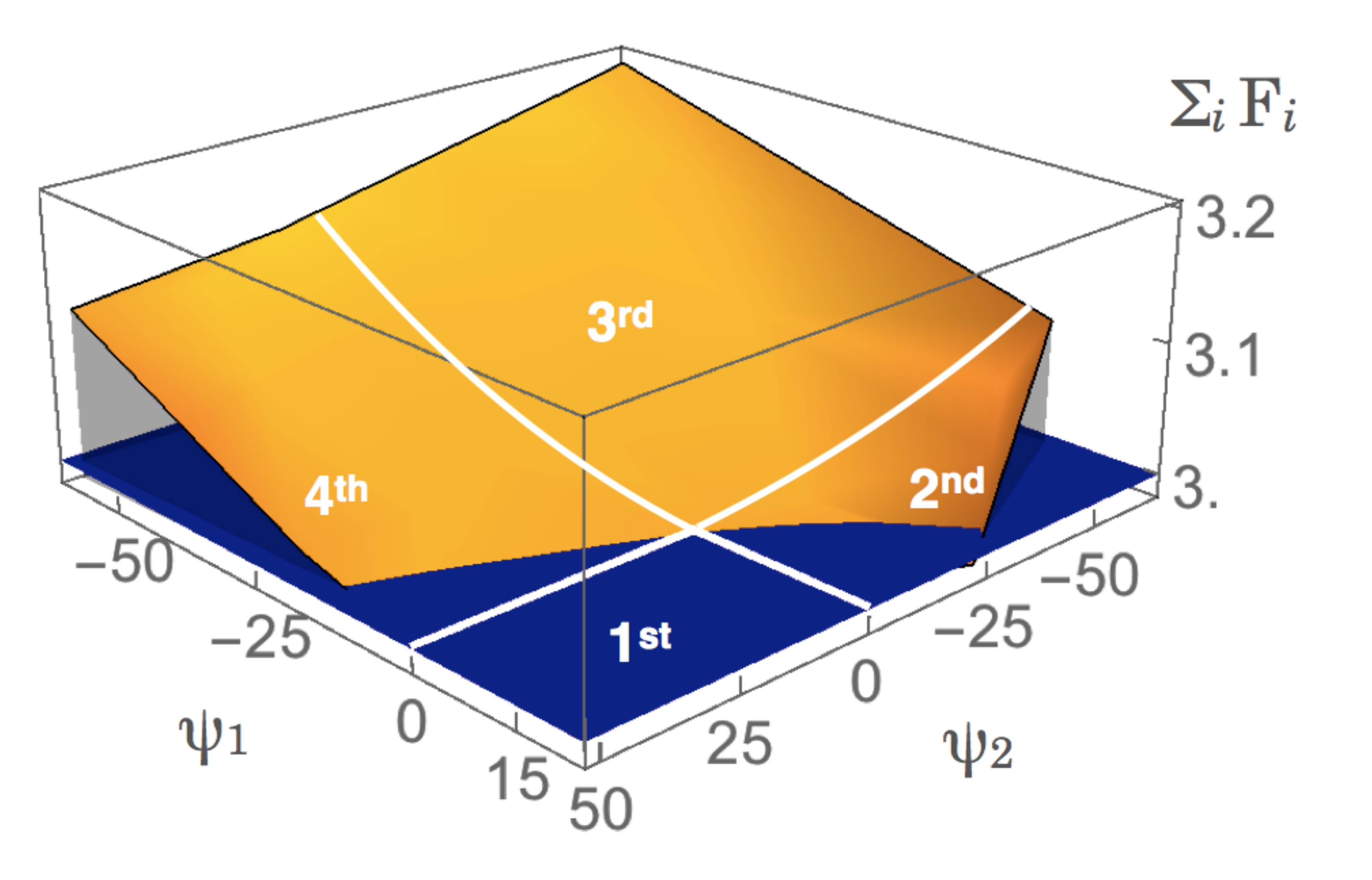}
	\caption{Sum of Fano factors over the three different species as a function of the two independent mean fluxes $\Psi_1$ and $\Psi_2$, for  different values of parameters $f$ and $g$, for the SDEs shown in Fig. \ref{fig1}b with $n$ and $m$ equal to 2 ; all other parameters are kept fixed: $p_x=p_y=p_z=100$, $d_x=d_y=d_z=0.001$. The blue surface is the surface $\sum_i F_i = 3$ while the orange is obtained by solving analytically the SDE systems in the moment closure approximation. In the quadrants I and III,  when the signs of the fluxes are equal, there is, respectively,  an increase or a decrease of the global noise. In the other quadrants, both scenarios are possible, according to the parameters of the model.} 
\label{Fano2}
\end{figure}

We showed analytically and numerically that the global intrinsic noise in these two CRNs, defined as  the sum of the Fano factors over all species, is given by :
\begin{equation}
\sum_{i=1}^{\text{card}(\mathcal{S})} F_i =   \mathcal{X}  - \sum_{j=1}^\delta \alpha_j   (-\sum_i^{\text{card}(\mathcal{S})} k_{ij}) \, \Psi_j
\label{eq222}
 \end{equation}

\noindent 
where all $\alpha_j$ are positive functions of the parameters and where ($-\sum_i k_{ij}$) is  the reaction's stoichiometry, equal to $(n-1)$, $(m-1)$ and $(n m-1)$ for $j=1,2,3$, respectively; the rank is  $\mathcal{X}= 3$. The sum is over the $\delta$ independent internal mean fluxes expressed as a function of forward and backward reactions as $\Psi_j=\textbf{E}[a_j-a_{-j}]$, for which the stoichiometry of reactant and product differ.  The analytical results for   system (\ref{CRN2}) are plotted in Fig. \ref{Fano2}. 

Eq. (\ref{eq222}) shows that when all the  fluxes flow from the species of lower degree of complexity to the species of higher complexity, the global intrinsic noise level is reduced. In contrast, when the fluxes are directed towards
the lower complexity species, it is amplified. Again, the effect on the noise  is higher for larger stoichiometry values ($-\sum_i k_{ij}$). When the fluxes do not all have the same sign, there start to be a competition between the fluxes. In that case, whether the noise is reduced or amplified depends on the parameters of the system, as pictorially shown in Fig.
\ref{Fano2}. 

%\\begin{widetext}
%\\begin{align}
%\dX(t) = dP_X(t) - dD_X(t) - n\, (dF(t)- dG(t)) - dL(t) + dM(t) \nonumber \\
%\dY(t) = dP_Y(t) - dD_Y(t) - n\, (dH(t)- dI(t)) \,+ dL(t)- dM(t)  \nonumber \\
%\dZ(t) = dP_Z(t) - dD_Z(t) + (dF(t)- dG(t) + dH(t)- dI(t))
%\\end{align}
%\\end{widetext}

%\where production and degradation terms are similar to those in \ref{oligomerizationII} while the interconvertion terms are given by

%\\begin{eqnarray}
%\dF(t)  = f X(t)^n dt +  \sqrt{f X(t)^n}\, \Gamma_{f} (t) \nonumber \\
%\dG(t) = g Z(t) dt + \sqrt{g Z(t)}\, \Gamma_{g} (t) \nonumber \\
%\dH(t)  = h Y(t)^n dt +  \sqrt{h Y(t)^n}\, \Gamma_{h} (t) \nonumber \\
%\dI(t) = i Z(t) dt + \sqrt{i Z(t)}\, \Gamma_{i} (t) \nonumber \\
%\dL(t)  = l X(t) dt +  \sqrt{l X(t)}\, \Gamma_{l} (t) \nonumber \\
%\dM(t) = m Y(t) dt + \sqrt{m Y(t)}\, \Gamma_{m} (t) \nonumber \\
%\\label{oligomerization5}
%\\end{eqnarray}

\section{Conclusion}

In this letter we analyzed the global intrinsic noise in CRNs through the estimation of the sum of Fano factors over all chemical species involved. We found that this quantity depend crucially on the value of the CRN's deficiency.  For all weakly reversible CRNs with $\delta$ = 0 the sum of Fano factors is always constant and equal to the rank of the system independently of the model's parameter. For higher deficiency systems, additional terms appear which are proportional to the fluxes between the complexes times the reaction stoichiometry. If all fluxes flow in the direction of higher complexity, a global reduction of the noise is observed, while an amplification occurs when fluxes are directed towards lower complexity.   

To get insights into the biological meaning of our results, consider for example the system composed of monomeric proteins that undergo an homo-oligomerization process. In this case, that corresponds to Fig. \ref{fig1}a with $p_y=0$, the mean flux is directed towards higher complexity and thus the sum of Fano factors is always smaller than or equal to the rank, which  signals global noise reduction. In contrast, for systems modeling the chemical hydrolysis of cellulose and hemicellulose into monomeric sugars, where instead $p_x=0$ and the flux is directed towards the lowest complexity species, we always have global noise amplification. 

Several points remain to be addressed. The analysis of the Fano factors of each individual chemical species in the CRN are left for an upcoming paper \cite{PucciRoomanII}.  We will also extend our results to systems with more general kinetic schemes \cite{Seifert,Bustamante,AndersonGeneralized}.  Last but not least, a clear interpretation of our results in terms of entropy production rates will contribute to deepen our physical understanding of the noise modulation in CRNs \cite{Polettini,Rao}.

\begin{acknowledgments}
FP is Postdoctoral researcher and MR Research Director at the Belgian Fund for Scientific Research (FNRS). 
\end{acknowledgments}

\end{document}